\documentclass{article}

    \usepackage[final]{neurips_2022}

\usepackage[utf8]{inputenc} % allow utf-8 input
\usepackage[T1]{fontenc}    % use 8-bit T1 fonts
\usepackage{hyperref}       % hyperlinks
\usepackage{url}            % simple URL typesetting
\usepackage{booktabs}       % professional-quality tables
\usepackage{amsfonts}       % blackboard math symbols
\usepackage{nicefrac}       % compact symbols for 1/2, etc.
\usepackage{microtype}      % microtypography
\usepackage{xcolor}         % colors

\usepackage{amsmath}
\usepackage{amssymb}
\usepackage{mathtools}
\usepackage{amsthm}
\usepackage{subcaption}
\usepackage{amsmath, amsfonts, amssymb}
\usepackage{array}
\usepackage{subcaption}
\usepackage{xfrac}

\title{PhysGNN: A Physics--Driven Graph Neural Network Based Model for Predicting Soft Tissue Deformation in Image--Guided Neurosurgery}

\author{%
  Yasmin Salehi, \hspace{2pt}Dennis Giannacopoulos \\
  Department of Electrical and Computer Engineering\\
  McGill University\\
  Montreal, QC, Canada \\
  \texttt{yasmin.salehi@mail.mcgill.ca, dennis.giannacopoulos@mcgill.ca} \\
}

\begin{document}

\maketitle

\begin{abstract}
Correctly capturing intraoperative brain shift in image-guided neurosurgical procedures is a critical task for aligning preoperative data with intraoperative geometry for ensuring accurate surgical navigation. While the finite element method (FEM) is a proven technique to effectively approximate soft tissue deformation through biomechanical formulations, their degree of success boils down to a trade-off between accuracy and speed. To circumvent this problem, the most recent works in this domain have proposed leveraging data-driven models obtained by training various machine learning algorithms---e.g., random forests, artificial neural networks (ANNs)---with the results of finite element analysis (FEA) to speed up tissue deformation approximations by prediction. These methods, however, do not account for the structure of the finite element (FE) mesh during training that provides information on node connectivities as well as the distance between them, which can aid with approximating tissue deformation based on the proximity of force load points with the rest of the mesh nodes. Therefore, this work proposes a novel framework, PhysGNN, a data-driven model that approximates the solution of the FEM by leveraging graph neural networks (GNNs), which are capable of accounting for the mesh structural information and inductive learning over unstructured grids and complex topological structures. Empirically, we demonstrate that the proposed architecture, PhysGNN, promises accurate and fast soft tissue deformation approximations, and is competitive with the state-of-the-art (SOTA) algorithms while promising enhanced computational feasibility, therefore suitable for neurosurgical settings.

\end{abstract}

\section{Introduction}
Exploiting high-quality preoperative images for surgical planning is a common practice among neurosurgeons. With the recent advances in surgical tools, medical imagery techniques have especially made their way into operating rooms to enhance surgical visualization and navigation, giving rise to image-guided neurosurgical techniques, where correctly aligning patient's intraoperative anatomical geometry with preoperative images has a significant impact on the accuracy of surgical navigation \citep{hagemann1999biomechanical, warfield2000real, ferrant2002serial, sun2004modeling, vigneron2004modelling, choi2004efficient, miller2007total, wittek2009unimportance, joldes2009suite, vigneron2010enhanced, wittek2010patient, li2014framework, miller2019biomechanical}. This task, however, poses a challenge. During the neurosurgical procedure, the brain continuously undergoes a series of anatomical changes for a variety of reasons such as gravity, craniotomy,\footnote{Temporarily removing part of the bone from skull to access the brain tissue.} tissue retraction and resection, and anesthetics \citep{roberts1998intraoperative}, making surgical navigation based on preoperative data alone error-prone and unreliable \citep{wittek2010patient}. 

To account for intraoperative brain shift,\footnote{Volumetric brain deformations induced by surgical operations \citep{nabavi2001serial}.} registration technology, by which the coordinate system of preoperative data and surgical site are brought into spatial alignment, has emerged as an essential tool in image-guided surgical frameworks for maintaining surgical navigation accuracy and depth perception \citep{crum2004non, cleary2010image, risholm2011multimodal}. In a registration framework, data alignment is achieved by leveraging interpolation and transformation functions, which manipulate preoperative data to realize tissue deformations and are regularized by a constraint \citep{ferrant2002serial}. 

At large, registration approaches may be categorized into image-based and model-based methods, which differ in the type of prior information they use as a constraint for deforming preoperative data with respect to intraoperative geometry. To define, image-based registration methods are a set of techniques which mainly use pixel/voxel information and solely rely on image processing techniques to realize tissue deformation \citep{beauchemin1995computation, hata2000three, ji2014cortical, iversen2018automatic, tu2019survey}. While they may be effective in some settings---e.g., registering deformations of hard tissues such as bone---they are prone to failing in effectively capturing \textit{soft} tissue deformation for various reasons, such as contrast variations between preoperative and intraoperative images resulted from use of different image acquisition techniques, disappearance of boundaries between tissues upon surgical cuts, and presence of contrast agents \citep{ferrant2002serial}. But most importantly, image-based techniques neglect prior information about the biomechanical behaviour of soft tissue, which is an essential knowledge for creating accurate volumetric mappings \citep{ nabavi2001serial, ferrant2002serial}.

On the other hand, model-based methods are a set of registration techniques that treat images as a \textit{deformable volume}---a notion first introduced by \citet{broit1981optimal}---to better allow for presenting elastic and plastic deformations. These approaches address intraoperative registration challenges by imposing additional constraints---based on mathematical or physical characterization of soft tissue \citep{zhang2017deformable}---to compute soft tissue deformation, resulting in mathematical- and physics-based solution formulations. In particular, studying physics-based methods has become an emerging area of research in recent years \citep{choi2004efficient, joldes2009suite, malone2010simulation, wittek2010patient, joldes2010real, basafa2011real, miller2019biomechanical} for their potential to accurately model the biomechanics of the brain and \textit{predicting} soft tissue deformation. Indeed, recent advances in this domain have led to the innovation of advanced image-guided surgical systems that are equipped with augmented reality (AR), which is established by fusing virtual reality (VR) with high-quality preoperative images in the form of a deformable volume \citep{cleary2010image}. The goal of integrating AR in this framework is to facilitate tracking predefined surgical access routes without having to look away from the screen, and provide surgeons with visual tool-tissue interaction \citep{tonutti2017machine}. 

In the literature, the finite element method (FEM) is a powerful technique for generating 3-dimensional (3D) deformable volumes from high-quality preoperative images, which can simulate intraoperative brain deformations \citep{wittek2010patient}. However, their degree of success boils down to a trade-off between accuracy and speed \citep{tonutti2017machine}. In most cases, benefiting from a finite element (FE) model, which accounts for all types of material and geometric non-linearities, is not always feasible for requiring significant computational resources \citep{wittek2010patient, tonutti2017machine}. To address this limitation, recent works have proposed deriving deformable models by training machine learning algorithms with the results of precomputed FEM simulations that are based on preoperative images \citep{de2011physics, tonutti2017machine, lorente2017framework, pfeiffer2019learning, liu2020real}. However, while promising, they all lack a key aspect for effectively learning over mesh data: they do not account for graph structure---e.g., node connectivities---which is an important piece of information for learning nodal displacements within each FE resulted from prescribed loads. In fact, wanting to incorporate node connectivities by feeding the adjacency matrix as a set of features to the aforementioned algorithms will result in an ill-posed problem as the number of features would then exceed the number of data points. Additionally, these methods are arguably computationally inefficient and hard to train for high quality meshes---which contain many nodes---as the number of parameters needed to train the models would be in order of $O(|\mathcal{V}|)$, where $|\mathcal{V}|$ is the number of mesh nodes. Although the later problem has been previously addressed by training only on a subset of the mesh data \citep{tonutti2017machine}, or capturing a lower dimensional embedding of the deformation space through Principle Component Analysis (PCA) \citep{liu2020real}, an important question is raised on whether we can enhance computational feasibility while also benefiting from the structural information of the FE mesh.

\paragraph{Present Work} In this work, we propose a novel data-driven framework, named PhysGNN,\footnote{The data and implementation are available at \url{https://github.com/YasminSalehi/PhysGNN}.} which approximates soft tissue deformation under prescribed forces by leveraging graph neural networks (GNNs) \citep{scarselli2008graph}. Specifically, PhysGNN is a physics-driven framework for having been trained with the results of Finite Element Analysis (FEA)---that is based on the physical properties of soft tissue such as Young's modulus and Poisson ratio---and for approximating tissue deformation---i.e., displacement of the FE mesh nodes---by summing up the forces applied to the prescribed load nodes and their neighbours using GraphSAGE \citep{hamilton2017inductive} and GraphConv \citep{morris2019weisfeiler} layers, analogous to the way displacement caused by applied force(s) is computed according to the Newton's first law of motion. To the best of our knowledge, we are the first ones to be leveraging GNNs for approximating soft tissue behavior upon prescribed loads, and propose combination of GraphSAGE and GraphConv layers as well as Jumping Knowledge (JK) connections \citep{xu2018representation} for this purpose. Our motivation for using GNNs for this task is their ability to enable generalization of convolution on graph data through the notion of \textit{neural message passing} \citep{gilmer2017neural}, by which the information of nodes within a neighbourhood of an arbitrary size are transformed and aggregated producing a new value \citep{hamilton2020graph}. The message passing framework enables inductive learning over graphs through \textit{parameter sharing}, a property that ensures learning a constant number of parameters independent of the mesh size. Therefore, contrary to the previous methods, not only is PhysGNN capable of incorporating mesh structural information---i.e., proximity of mesh nodes with each other and the Euclidean distance of neighbouring nodes from one another---which can aid with effective prediction of tissue deformation, but also can enhance computational feasibility when learning from high quality meshes that consist of many nodes.

We evaluate the performance of various PhysGNN models---resulted from different combination of GraphSAGE and GraphConv layers and aggregation functions---on two datasets---distinguished by the number of force load points and the total force applied---against the results of FEA (our baseline). Empirically, we demonstrate that PhysGNN is capable of successfully predicting tissue deformation upon prescibed forces under 0.005 and 1.9 seconds using a GPU and CPU, respectively. In fact, the absolute displacement error of the best-performing PhysGNN model for 97\% of the nodes was found to be under 1 mm---which is the precision in neurosurgery \citep{miller2010biomechanics}.

\section{Related Work}

\begin{figure}[t]
\centering
\includegraphics[width=0.98\columnwidth]{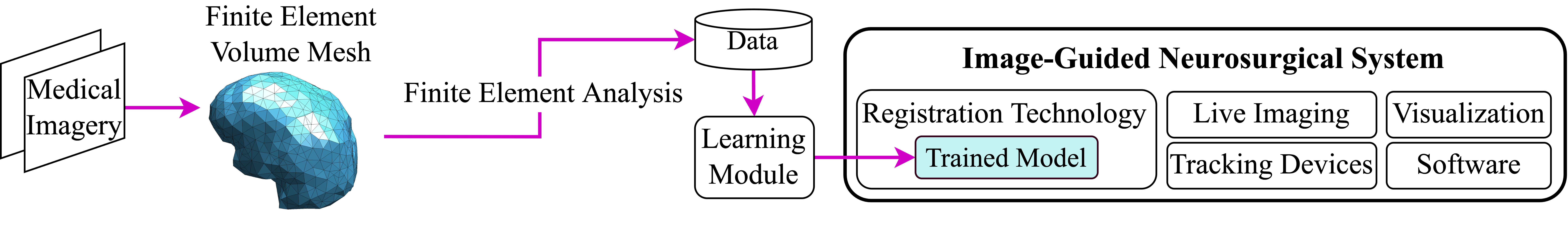}
\caption{A generic image-guided neurosurgical system framework \citep{cleary2010image}, where patient's anatomy is aligned with preoperative images through deforming a deformable volume generated from preoperative imagery, which its biomechanical behavior under prescribed loads is predicted by machine learning methods.}
\label{3d_registration}
\end{figure}

\paragraph{Approximating Tissue Deformation with Machine Learning} The predictive power of various machine learning algorithms in approximating tissue deformation with respect to physics laws has been assessed in a number of studies. These approaches either implement machine learning techniques as an alternative to the FEM---e.g., representing mass points as cellular neural networks to assimilate elastic deformations \citep{zhong2006cellular, zhang2019neural}---or as a subsequent step to preoperative FEA in which data-driven models are resulted by training machine learning models with patient-generated content that are intrasubject (patient-specific) to predict soft tissue deformation, as shown in Figure \ref{3d_registration}. For example, while \citet{de2011physics} proposed a physics-driven framework, named PhyNNeSS, consisting of radial basis function to reconstruct the deformation fields of liver and Penrose drain models, \citet{lorente2017framework} investigated the success of linear regression, decision trees, random forests, and extremely randomized trees in simulating the biomechanical behavior of human liver under large deformations and during breathing in real time. Furthermore, \citet{liu2020real} exploited ANNs to reconstruct tissue deformation of a head and neck tumour, where the dataset was encoded into a low-dimensional representation using PCA prior to training to facilitate the learning process. Meanwhile, similar techniques have been employed to approximate tissue deformation in neurosurgical settings. In \citep{tonutti2017machine}, the authors explored and compared the effectiveness of ANNs and support vector regression (SVR) in predicting deformation of a brain tumour caused by applying forces to the brain tissue. Generalizing the notion of deep learning based soft tissue simulation frameworks to different types of organs, \citet{pfeiffer2019learning} proposed a convolutional neural network (CNN) based soft tissue simulation system to predict an organ’s internal deformation from known surface deformation. In their study, they accounted for \textit{all} tissue types within an organ and trained CNNs on an entirely synthetic data of random organ-like meshes to enhance the learning process by using more data than otherwise available. They stated their method can be adopted by the neurosurgical community to capture brain shift, or used in motion compensation radio-therapy. 

\paragraph{Learning Mesh-Based Simulations Using GNNs} Recently, \citet{sanchez2020learning} and \citet{pfaff2020learning} have proposed Graph Network-based Simulators (GNS) and MeshGraphNets, respectively, to simulate interaction of fluids, rigid solids, and deformable materials with one another \citep{sanchez2020learning}, and complex physical systems such as aerodynamics, structural mechanics, and cloth \citep{pfaff2020learning}. Both GNS and MeshGraphNets are generative architectures that use GraphNet \citep{sanchez2018graph} as their building block. 

% \noindent\textbf{Learning Mesh-Based Simulations Using GNNs.} Recently, \citet{sanchez2020learning} and \citet{pfaff2020learning} have proposed Graph Network-based Simulators (GNS) and MeshGraphNets respectively---both GraphNet \citep{sanchez2018graph} based generative architectures---to simulate interaction of fluids, rigid solids, and deformable materials with one another \citep{sanchez2020learning}, and complex physical systems such as aerodynamics, structural mechanics, and cloth \citep{pfaff2020learning}.

% \citet{belbute2020combining} have previously proposed a Graph Convolution Network (GCN) \citep{kipf2016semi} based architecture to learn fluid flow prediction.

\section{The Physics-Driven Graph Neural Network (PhysGNN) Model}

PhysGNN is a data-driven model that leverages GNNs to learn the nonlinear behavior of soft tissue under prescribed loads from FE simulations. In this work, GNNs are deemed as the most suitable deep learning candidates for capturing tissue deformation for the reasons stated below. 
\begin{enumerate}

\item First, GNNs enable incorporating mesh structural information by forming a computation graph---a neural network architecture---for every node through unfolding the neighbourhood around them via neural message passing, which refers to the exchange of feature and structural information between nodes that are within the same $k$-hop neighbourhood \citep{hamilton2020graph}. This characteristic of GNNs can ensure more accurate prediction of tissue deformation that is especially in close proximity of prescribed forces and susceptible to undergoing larger displacements, since the distance between prescribed force nodes and free-boundary condition nodes can be accounted for. 

\item Second, GNNs learn the same weight and bias parameters across the FE mesh through parameter sharing. In other words, the number of parameters which need to be learned are constant and independent of the mesh size, a property that also enables generalization of GNNs to unseen parts of the graph. This makes training PhysGNN with high quality meshes---which inherently consist of many nodes---computationally less expensive than the methods presented in the previous studies.
\end{enumerate}

% as illustrated in Figure \ref{fig:physgnn_micro}.

\subsection{Architecture}

% PhysGNN is a data-driven model by which the deformation is captured microscopically---i.e., the displacement of mesh nodes is approximated by summing up forces applied to the nodes within their $k$-hop neighbourhood and/or propagating information of the nodes to which the maximum force has been applied at each $k$-hop for K iterations where $k \in \{1,...,K\}$. 

PhysGNN is a data-driven model that captures tissue deformation---i.e., displacement of the FE mesh nodes---caused by prescribed forces microscopically by summing up or taking the maximum of forces applied to their $k$-hop neighbours at each $k$ for K iterations, where $k \in \{1,...,K\}$. The idea of force summation---or taking the maximum for when force is applied to only one of the mesh nodes---is based on the way net displacement due to a force is computed according to Newton’s first law of motion, and inspired by the force propagation model presented in \citep{choi2004deformable}. For this purpose, PhysGNN incorporates two types of GNNs, namely GraphSAGE \citep{hamilton2017inductive} and GraphConv---the graph convolution operator of \emph{k}-GNNs in \citep{morris2019weisfeiler}---where the later GNN allows for incorporating nodal distances. The message passing architecture of GraphSAGE and GraphConv for a graph $\mathcal{G}= (\mathcal{V,E})$, with $\mathcal{V}$ and $\mathcal{E}$ being the vertex and edge sets, respectively, $u,v \in \mathcal{V}$, and matrix $\boldsymbol{X} \in \mathbb{R}^{d \times |\mathcal{V}|}$ representing the node features,
are defined as:
% \begin{align}
% \mathbf{m}_{\mathcal{N}(u)}^{(k)} &=  \textsc{Aggregate}^{(k)}(\{\textbf{h}_v^{(k-1)}, \forall v \in \mathcal{N}(u)\}) \notag \\
% \mathbf{h}_u^{(k)} &= \big[\mathbf{W}_{\text{self}}^{(k)} \cdot {\textbf{h}}_u^{(k-1)} \oplus \mathbf{W}_{\text{neigh}}^{(k)} \cdot\mathbf{m}_{\mathcal{N}(u)}^{(k)}
%  + \mathbf{b}^{(k)}  \big],
% \end{align} 
% and 
% \begin{align}
% \mathbf{m}_{\mathcal{N}(u)}^{(k)} &=  \textsc{Aggregate}^{(k)}(\{e_{vu} \cdot\textbf{h}_v^{(k-1)}, \forall v \in \mathcal{N}(u)\}) \notag \\
% \mathbf{h}_u^{(k)} &= \textsc{Update}^{(k)}\big[\mathbf{W}_{\text{self}}^{(k)} \cdot {\textbf{h}}_u^{(k-1)}, \mathbf{W}_{\text{neigh}}^{(k)} \cdot\mathbf{m}_{\mathcal{N}(u)}^{(k)}
%  + \mathbf{b}^{(k)}  \big],
% \end{align}
\begin{align}
\mathbf{m}_{\mathcal{N}(u)}^{(k)} &=  \textsc{Aggregate}^{(k)}(\{\textbf{h}_v^{(k-1)}, \forall v \in \mathcal{N}(u)\}) \notag \\
\mathbf{h}_u^{(k)} &= \sigma \bigg(\mathbf{W}_{u}^{(k)} \cdot {\textbf{h}}_u^{(k-1)} \oplus \mathbf{W}_{\mathcal{N}(u)}^{(k)} \cdot\mathbf{m}_{\mathcal{N}(u)}^{(k)}
 + \mathbf{b}^{(k)}  \bigg),
\end{align} 
and 
\begin{align}
\mathbf{m}_{\mathcal{N}(u)}^{(k)} &=  \textsc{Aggregate}^{(k)}(\{e_{vu} \cdot\textbf{h}_v^{(k-1)}, \forall v \in \mathcal{N}(u)\}) \notag \\
\mathbf{h}_u^{(k)} &= \sigma \bigg(\textsc{Merge}^{(k)}\big(\mathbf{W}_{u}^{(k)} \cdot {\textbf{h}}_u^{(k-1)}, \mathbf{W}_{\mathcal{N}(u)}^{(k)} \cdot\mathbf{m}_{\mathcal{N}(u)}^{(k)} \big)
 + \mathbf{b}^{(k)}  \bigg),
\end{align}
respectively, where $\mathcal{N}(u)$ is the neighbourhood of node $u$, $m_{\mathcal{N}(u)}^{(k)}$ represents the aggregated messages from the neighbours of node $u$ at iteration $k$, where $k \in \{1, \dots, K\}$, $\mathbf{h}_u$ and $\mathbf{h}_v$ are the embedding for nodes $u$ and $v$, respectively,  $e_{uv}$ is the edge weight between node $u$ and $v$, $\oplus$ is the concatenation operator, \textsc{Aggregate} is an arbitrary permutation-invariant, differentiable function, \textsc{Merge} is either a sum, a column-wise vector concatenation or a long short-term memory (LSTM) style update step, $\sigma$ is a component-wise non-linear function (e.g., a sigmoid, or a ReLU), and the weights, $ \mathbf{W}_{u}^{(k)},\mathbf{W}_{\mathcal{N}(u)}^{(k)} \in \mathbb{R}^{d^k \times d^{(k-1)}}$, and the bias term, $\mathbf{b}^{(k)} \in \mathbb{R}^{d^{(k)}}$, are trainable parameters. Moreover, the input to the GNN is simply the node features, expressed as $
\mathbf{h}_u^0 = \mathbf{x}_u , \forall u \in \mathcal{V}$, while the output of the final layer for each node is $\mathbf{z}_u =\boldsymbol{h}_u^K, \forall u \in \mathcal{V}$ \citep{hamilton2017inductive, morris2019weisfeiler, hamilton2020graph}.

\begin{figure}[t]
\centering
\includegraphics[width=0.98\textwidth]{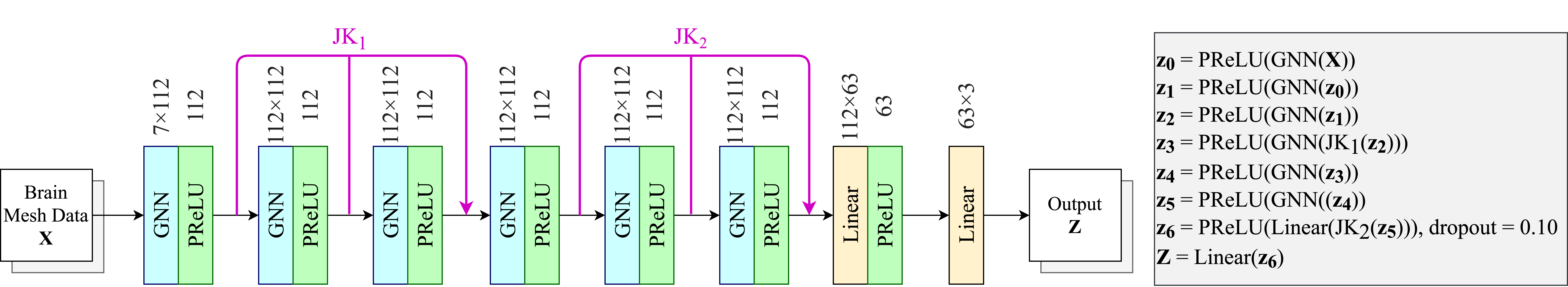}
\caption{Architectural diagram of PhysGNN where the "GNN" layers may be GraphConv and/or GraphSAGE layers that aggregate messages by summation and/or maximum functions. The best-performing PhysGNN models in our study are shown in Figure \ref{fig:best_config_1} and \ref{fig:best_config_2}.}
\label{PhysGNN_arch}
\end{figure}

In this work, $K$ is set to 6---i.e., the mesh nodes capture information from their 6-hop neighborhood. This decision was made by noting performance improvement after increasing the number of GNN layers. Although a deeper neural network architecture may have yielded a better performance, increasing the number of GNN layers beyond 6 increased training time while may have also resulted in over-smoothing for aggregating over the entire FE mesh. Moreover, the implemented \textsc{Aggregate} functions are addition and maximum for summing up or taking the maximum of the applied forces, respectively, while the \textsc{Merge} function used for the GraphConv layers is concatenation. PhysGNN also benefits from a variation of JK connections \citep{xu2018representation} to combat over-smoothing, which generates a weighted sum of embeddings at different iterations from a bi-directional LSTM layer as:
\begin{equation}
\sum_{t=1}^T\mathbf{\alpha}_u^{(t)} {\textbf{h}}_u^{(t)},
\end{equation}
where $\mathbf{\alpha}_u^{(t)}$ are the attention scores. A full architectural diagram of PhysGNN is depicted in Figure \ref{PhysGNN_arch}.  

% as $\sum_{t=1}^T\mathbf{\alpha}_u^{(t)} {\textbf{h}}_u^{(t)}$, where $\mathbf{\alpha}_u^{(t)}$ are the attention scores. A full architectural diagram of PhysGNN is illustrated in Figure \ref{PhysGNN_arch}. 

It should be noted that different PhysGNN models can be obtained by varying the combination of GraphConv and GraphSAGE layers and aggregation functions. The performance of various PhysGNN models on two different datasets is presented in the Results and Discussion section. 

\section{Experimental Setting}

\subsection{Data Acquisition }

\begin{figure}[t]
\centering
\begin{subfigure}{0.328\textwidth}
  \includegraphics[width=\linewidth]{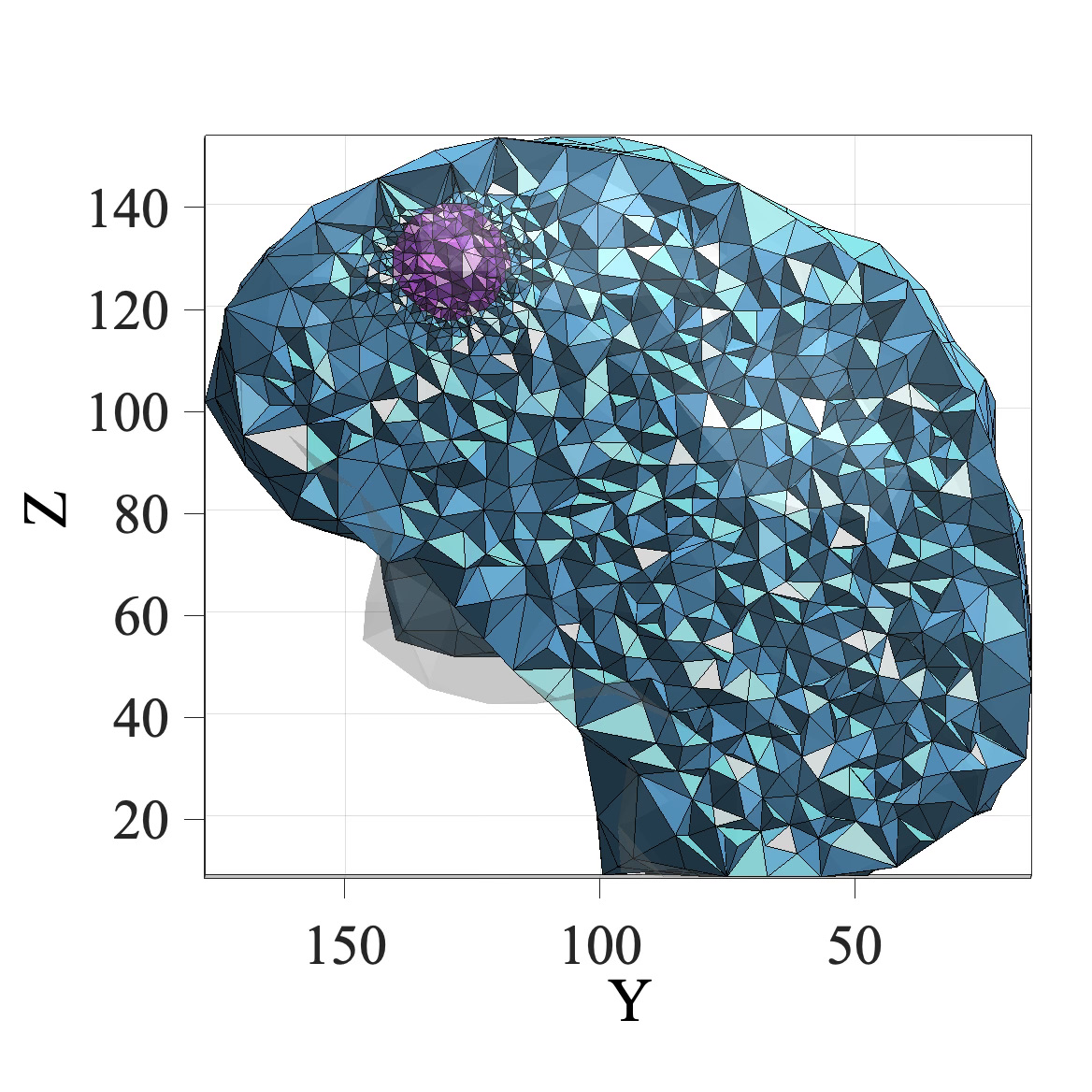}
  \captionsetup{justification=centering}
  \caption{}
  \label{fig:FEmesh}
\end{subfigure} 
\begin{subfigure}{0.328\textwidth}
  \includegraphics[width=\linewidth]{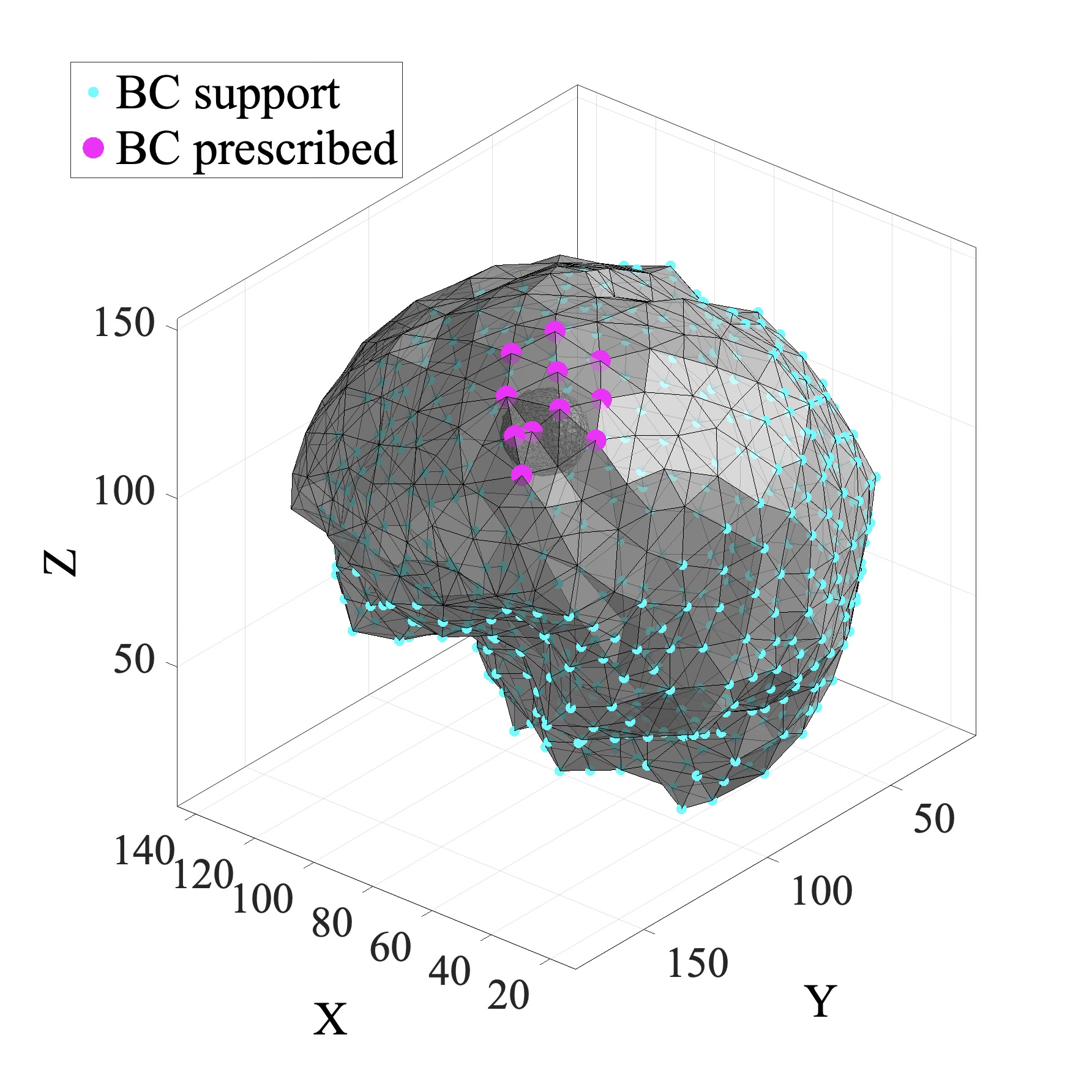}
  \captionsetup{justification=centering}
  \caption{}
  \label{fig:dataset1}
\end{subfigure}
\begin{subfigure}{0.328\textwidth}
  \includegraphics[width=\linewidth]{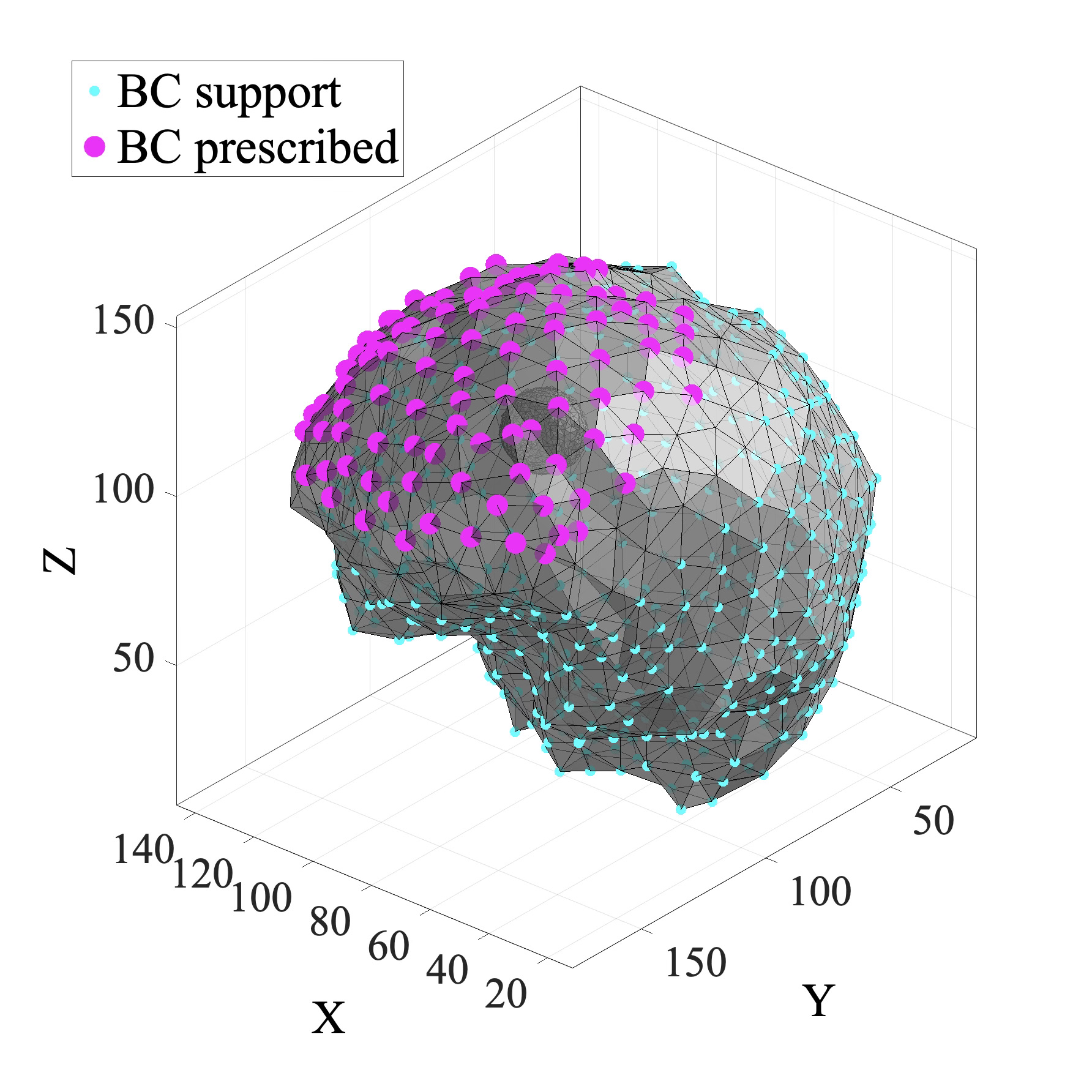}
  \captionsetup{justification=centering}
  \caption{}
  \label{fig:dataset2}
\end{subfigure}
\caption{(a) Cut view of the FE volume mesh, where cyan represents the healthy brain tissue, while magenta represents the tumour tissue; prescribed and fixed boundary condition nodes, denoted as BC prescribed and BC support, respectively, of (b) Dataset 1 and (c) Dataset 2.}
\label{fig:datasets}
\end{figure}

\paragraph{Generating the Finite Element Mesh} The FE volume mesh for running FE simulations was generated by running TetGen \citep{si2015tetgen} on the surface mesh files provided by \citet{tonutti2017machine}, which is based on MRI scans taken from the Repository of Molecular Brain Neoplasia Data \citep{madhavan2009rembrandt}. Figure \ref{fig:FEmesh} is illustrative of the FE volume mesh, which consists of 9118 nodes. More details on the FE volume mesh may be found in Table \ref{tab:mesh_info} in the appendix.

\begin{table}[t]
\caption{Material properties of different parts of the brain \citep{joldes2010real, wittek2010patient}.}
    \centering
    \small
    \begin{tabular}
    {>{\centering\arraybackslash}m{0.85in}
    >{\centering\arraybackslash}m{0.5in}
    >{\centering\arraybackslash}m{0.8in}
    >{\centering\arraybackslash}m{0.5in}
    >{\centering\arraybackslash}m{0.5in}
    >{\centering\arraybackslash}m{0.5in}
    >{\centering\arraybackslash}m{0.5in}
    }
    \toprule  
     Dataset & Density (kg/$\text{m}^3)$ & Young's Modulus ($\text{Pa}$) & Poisson Ratio  \\
    \toprule
    Healthy Tissue & 1000 & 3000 & 0.49\\ \hline
    Tumour Tissue & 1000 & 7500 & 0.49 \\
    \toprule
    \end{tabular}
    \label{tab:mat_properties}
\end{table}

\paragraph{Material Properties} Based on the work done in \citep{joldes2010real} and \citep{wittek2010patient}, the mechanical behavior of healthy brain and tumour tissues were characterized as hyper-elastic material obeying the Neo-Hookean constitutive law \citep{maas2012febio}. Additionally, similar to the experimental setting in \citep{tonutti2017machine}, the tumour and healthy brain tissues have identical density and Poisson ratio, and the tumour is assumed to be stiffer than the healthy brain tissue, thus having a larger Young's modulus. The numerical value of material parameters are listed in Table \ref{tab:mat_properties}.

% \paragraph{Material Properties} Based on the work done in \citep{joldes2010real} and \citep{wittek2010patient}, the mechanical behavior of healthy brain and tumour tissues were characterized as hyper-elastic material obeying the Neo-Hookean constitutive law, $W = \frac{2\mu}{\alpha^2}(\lambda_1^\alpha+\lambda_2^\alpha+\lambda_3^\alpha-3)$, where $W$ is the energy, $\mu$ is the shear modulus in undeformed state, $\alpha$ is a material coefficient, and $\lambda$ is the principal longitudinal stretch. Similar to the experimental setting in \citep{tonutti2017machine}, the tumour and healthy brain tissues have identical density and Poisson ratio, and the tumour is assumed to be stiffer than the healthy brain tissue, thus having a larger Young's modulus. The numerical value of material parameters can be found in Table \ref{tab:mat_properties}.

% \begin{equation}
% \small
%     W = \frac{2\mu}{\alpha^2}(\lambda_1^\alpha+\lambda_2^\alpha+\lambda_3^\alpha-3),
% \end{equation}

\begin{table}[t]
\caption{Characteristics of Datasets 1 and 2. B.C. stands for Boundary Condition. Pres. stands for Prescribed.}
\label{tab:dataset_info}
\centering
% \begin{small}
    \begin{tabular}
    {>{\centering\arraybackslash}m{0.39in}
    >{\centering\arraybackslash}m{0.35in}
    >{\centering\arraybackslash}m{0.35in}
    >{\centering\arraybackslash}m{0.64in}
    >{\centering\arraybackslash}m{0.66in}
    >{\centering\arraybackslash}m{0.45in}
    >{\centering\arraybackslash}m{0.58in}
    >{\centering\arraybackslash}m{0.67in}
    }
    \toprule
    Dataset & No. Fixed B.C. Nodes  & No. Pres. Load Nodes & No. Pres. Load Node(s) per Simulation  & No. Force Magnitudes &   Max. Force Applied (N) & No. Directions & No. Simulations \\
    \toprule 
    1 & 502 & 11  & 1   & 30 & 1.35 & 15 & 4950 \\ 
    \hline
    2 & 502 & 100 & 100 & 30 & 20 & 165 & 4950 \\
    \toprule
    \end{tabular}
% \end{small}
\end{table}

\paragraph{Finite Element Simulations} To assess the effectiveness of PhysGNN in predicting brain shift imposed by forces during surgery and especially craniotomy---which is modeled by applying forces to the surface of the brain \citep{wittek2009unimportance, miller2019biomechanical}---two different datasets were generated by running FE simulations on the FE mesh in FEBio \citep{maas2012febio} using the GIBBON \citep{moerman2018gibbon} toolbox. The datasets are distinguished by the total force applied and the number of prescribed load nodes. Specifically, Dataset 1 was created by applying 1.35 N to 1 node at a time in 30 time steps and 15 directions for a total of 11 different nodes. This choice was motivated by the results of the experimental study done by \citet{gan2015quantification} who measured 70\% of induced forces during neurosurgery to be less than 0.3 N and a maximal peak at 1.35 N during tissue dissection. On the other hand, Dataset 2 was created by applying 20 N to 100 nodes at a time in 30 time steps and 165 directions to further capture the non-linear behavior of soft tissue under large forces. Therefore, the amount of force applied to each prescribed node in Dataset 1 at time $i$ is:
\begin{equation}
    \mathbf{F}_i = \frac{\mathbf{F}_\text{total}}{30} \times i, \; i \in \{1,\dots,30\},
\end{equation}
whereas the amount of force applied to one of the prescribed nodes in Dataset 2 at time $i$ is:
\begin{equation}
    \mathbf{F}_i = \frac{\mathbf{F}_\text{total}}{30 \times 100} \times i, \; i \in \{1,\dots,30\}.
\end{equation}
The direction of forces applied to each prescribe load node features its surface normal direction, as well as directions that are randomly sampled from a hemisphere with radius 1. Lastly, to train PhysGNN, a train:validation:test split of 70:20:10 was applied to both datasets. Figures \ref{fig:dataset1} and \ref{fig:dataset2} are illustrative of the prescribed load nodes in Datasets 1 and 2, respectively, which are located near the tumour site, as well as the fixed boundary condition nodes, which mimic the skull bone that prevents tissue from moving. Further details on the generated datasets can be found in Table \ref{tab:dataset_info}. 

\begin{table}[t]
\caption{PhysGNN features and outputs.}
\label{tab:input_output}
\centering
\small
    \begin{tabular}{cc}
    \toprule
    Features ($\mathbf{X}$) & Output ($\mathbf{Z}$)\\
    \toprule
     $F_x, F_y, F_z, F_\rho, F_\theta, F_\phi,$ Physical Property & $\delta x$, $\delta y$, $\delta z$ \\
     \toprule 
    \end{tabular}
\end{table}

\subsection{Features and Output}
The features inputted into PhysGNN, are the force value applied to prescribed load nodes in Cartesian and spherical coordinates, denoted as $(F_x$, $F_y$, $F_z)$ and $(F_\rho, F_\theta, F_\phi)$, as well as a constant value, named Physical Property, which for nodes with free boundary condition is set to 0.4 if the node belongs to tumour tissue and 1 if the node belongs to healthy brain tissue, and 0 for nodes with fixed boundary condition. This value determines how much a node can undergo displacement based on its boundary condition and Young's modulus given that brain tissue is susceptible to undergoing larger displacements for its smaller Young's modulus compared to the tumour tissue. In particular, the value of 0.4 was calculated as the ratio of healthy tissue's Young's modulus (3000 Pa) over tumour tissue's Young's modulus (7500 Pa). On the other hand, the output of PhysGNN are the displacement value of the mesh nodes in the $x, y$, and $z$ directions, denoted as $\delta x$, $\delta y$, $\delta z$. Lastly, for the GraphConv models, the edge weights, $e_{u,v}$, were calculated as the inverse of the Euclidean distance between adjacent nodes $u$ and $v$ with $u,v \in \mathcal{V}$ as: \begin{equation}
e_{u,v} = \frac{1}{\sqrt{((x_u - x_v)^2 + (y_u - y_v)^2 + (z_u - z_v)^2)}}.    
\end{equation}
A summary of the features and output of PhysGNN can be found in Table \ref{tab:input_output}. 

% as:
% $\sfrac{1}{\sqrt{((x_u - x_v)^2 + (y_u - y_v)^2 + (z_u - z_v)^2)}}$.

\subsection{Baseline}

Similar to the previous studies \citep{tonutti2017machine, lorente2017framework, liu2020real}, the predictions made by the proposed data-driven model, namely PhysGNN in this work, are compared against the results approximated by the FEM.

\subsection{Loss Function, Optimization, and Regularization}
The loss function used for learning the trainable parameters is the mean Euclidean error computed as:
\begin{equation}
    \mathcal{L} = \frac{1}{|\mathcal{V}|}
    {\sum_{v \in \mathcal{V}}
    \sqrt{\sum_{i=1}^3
    (\mathbf{y}_{v,i}-\mathbf{z}_{v,i})^2}},
\end{equation}
where $|\mathcal{V}|$ is the number of mesh nodes, $\mathbf{y} \in \mathbb{R}^{|\mathcal{V}| \times 3}$ is the displacement in the $x$, $y$, and $z$ directions approximated by the FEM, and $\mathbf{z} \in \mathbb{R} ^{|\mathcal{V}| \times 3}$ is the displacement predicted by PhysGNN. To minimize the loss value, AdamW optimizer was used with an initial learning rate of 0.005, which was reduced by a factor of 0.1 to a minimum value of $1\mathrm{e}{-8}$ if no improvement on the validation loss was noticed after 5 epochs. To prevent over-fitting, early stopping was one of the regularization methods implemented by which learning stopped after no decrease in validation loss occurred after 15 epochs. Moreover, a dropout rate of 0.1 was applied to the second last layer of PhysGNN to improve generalization to unseen data.

\section{Results and Discussion} 

\subsection{Performance of PhysGNN in Predicting Tissue Deformation}
Table \ref{tab:results_data} is indicative of the performance of the PhysGNN models that resulted in the lowest validation loss on Datasets 1 and 2, illustrated in Figures \ref{fig:best_config_1} and \ref{fig:best_config_2}, respectively. According to the results, PhysGNN is capable of effectively predicting tissue deformation under prescribed loads, especially highlighted by
96\% and 97\% of absolute position errors being under 1 mm in Datasets 1 and 2, respectively. Notably, approximating tissue deformation using the FEM on a quad-core Intel i7 @ 2.9 GHz CPU took $11.5099 \pm 6.8004$ seconds on average for each FE simulation, with minimum and maximum computation times of $5.4191$ and $32.3798$ seconds, respectively, whereas prediction of tissue deformation per each FE simulation on a dual-core Intel(R) Xeon(R) @ 2.20GHz CPU took $1.8294 \pm 0.0322$ seconds on average, while on an NVIDIA P-100 GPU took $0.0043 \pm 0.0005$ seconds on average. Based on these observations and the fact that the accuracy in neurosurgery is not better than 1 mm \citep{miller2010biomechanics}, PhysGNN manifests itself as a promising deep learning module for predicting tissue deformation in neurosurgical settings for its accurate approximations, computation speed, and ease of implementation. Additional results on the median of errors may be found in Table \ref{tab:results_median} in the appendix. 

\begin{table}[t]
\caption{Best performance of PhysGNN on Datasets 1 and 2.}
\label{tab:results_data}
\centering
% \begin{small}
    \begin{tabular}
    {>{\centering\arraybackslash}m{0.57in}
    >{\centering\arraybackslash}m{0.52in}
    >{\centering\arraybackslash}m{0.52in}
    >{\centering\arraybackslash}m{0.52in}
    >{\centering\arraybackslash}m{0.56in}
    >{\centering\arraybackslash}m{0.45in}
    >{\centering\arraybackslash}m{0.55in}
    >{\centering\arraybackslash}m{0.46in}
    }
    \toprule
     Dataset & MAE ($\delta x$) (mm) & MAE ($\delta y$) (mm) & MAE ($\delta z$) (mm) & Mean Euclidean Error (mm) & Euclidean Error $ \leq 1\text{ mm}$ (\%)&
     Mean Absolute Position Error (mm) & Absolute Position Error$ \leq 1\text{ mm}$ (\%) \\
    \toprule
    Dataset 1 Validation & 0.1145 $\pm$ 0.2832 & 0.1051 $\pm$ 0.2544 & 0.1009 $\pm$ 0.2650 & 0.2105 $\pm$ 0.4530 & 94.99 & 0.1690 $\pm$ 0.4200 & 95.91 
    \\ \hline
    Dataset 1 Test & 0.1117 $\pm$ 0.2750 & 0.0999 $\pm$ 0.2318 & 0.0999 $\pm$ 0.2601 & 0.2049 $\pm$ 0.4329 & 95.11 & 0.1612 $\pm$ 0.3961 & 96.18
    \\ \toprule
    Dataset 2 Validation & 0.1478 $\pm$ 0.3825 & 0.1879 $\pm$ 0.5527 & 0.1283 $\pm$ 0.2364 & 0.3116 $\pm$ 0.6958 & 94.25 & 0.2106 $\pm$ 0.6129 & 96.26 \\
    \hline
    Dataset 2 Test & 0.1393 $\pm$ 0.3402 & 0.1842 $\pm$ 0.5479 & 0.1249 $\pm$ 0.2272 & 0.3023 $\pm$ 0.6671 & 94.60 & 0.2023 $\pm$ 0.5920 & 96.59\\
    \toprule
    \end{tabular}
% \end{small}
\end{table}

\begin{table}[t]
 \caption{The test set statistics of Datasets 1 and 2, where $\mathbf{y}$ is the displacement, and Max. Euclidean $\text{Error}_\text{mean}$ is computed as the average of the maximum Euclidean error observed for each data element---i.e., each simulation.}
\label{tab:max_vals}
\centering
% \begin{small}
    \begin{tabular}{cccc}
    \toprule
     Dataset & ${\delta \mathbf{y}}_{\text{max}}$ (mm) &  ${\delta \mathbf{y}}_{\text{mean}}$ (mm) & 
    Max. Euclidean $\text{Error}_\text{mean}$ (mm) \\
    \toprule
    1 & 24.5864 & 11.6229 $\pm$ 5.7681 & 2.5924 $\pm$ 1.5665 \\ 
    \hline
    2 & 47.8233 & 16.4170 $\pm$ 10.4650 & 4.1952 $\pm$ 2.5161 \\
    \toprule
    \end{tabular}
% \end{small}
\end{table}

\subsection{Comparison to Similar Studies}
Comparing our results with the most recent works in predicting tissue deformation with machine learning methods, our approach compares favourably. In \citep{tonutti2017machine} the authors investigated the effectiveness of their approach in approximating a brain tumour tissue deformation caused by applying forces less than 1 N to a single node on the brain surface and reported a mean absolute position error of $0.191 \pm 0.201$ mm. While our absolute position error for Dataset 1 (where forces $\leq$ 1.35 N are applied to a single node) seems to only be marginally smaller than theirs, we argue that our result is implicitly much better as our test set contains larger deformations since we consider for both the brain and tumour tissues. Similarly, our proposed method appears to be competitive with the results in \citep{lorente2017framework} where they reported 100\% of the Euclidean errors to be $\leq$ 1 mm, for a dataset with a maximum displacement of 15 mm. Lastly, \citet{liu2020real} reported an average Euclidean error of 0.129 mm and 0.392 mm, and an average maximum Euclidean error of 0.483 mm and 1.011 mm for two different test sets that exhibited a maximum displacement of 30 mm on a mesh with 1158 nodes. By referring to Tables \ref{tab:results_data} and \ref{tab:max_vals}, our results (achieved on a mesh with 9118 nodes) are shown to be competitive without requiring PCA to reduce the subspace dimensionality. Indeed, GNNs are inherently graph embedding models that encode graphs into a lower dimensional subspace to facilitate data manipulation for carrying out machine learning tasks on graphs. Therefore, we deem our method as not only effective for predicting tissue deformation but also a simpler one to implement.

\subsection{Ablation Study}

\begin{figure}[t]
\centering
 \begin{subfigure}[b]{0.9\textwidth}
    \includegraphics[width=\linewidth]{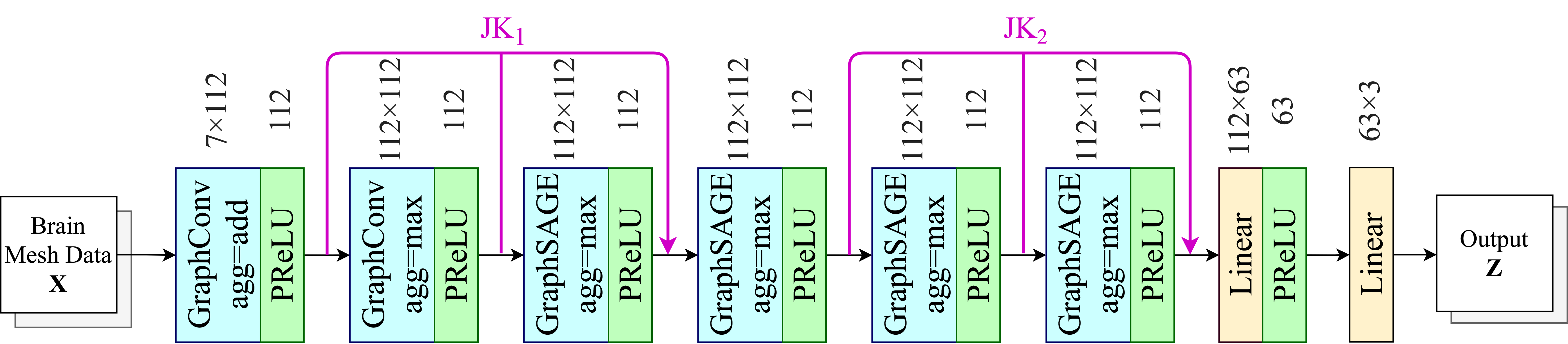}
    \captionsetup{justification=centering}
    \caption{}
    \label{fig:best_config_1}
    \end{subfigure}
    \begin{subfigure}[b]{0.9\textwidth}
    \includegraphics[width=\linewidth]{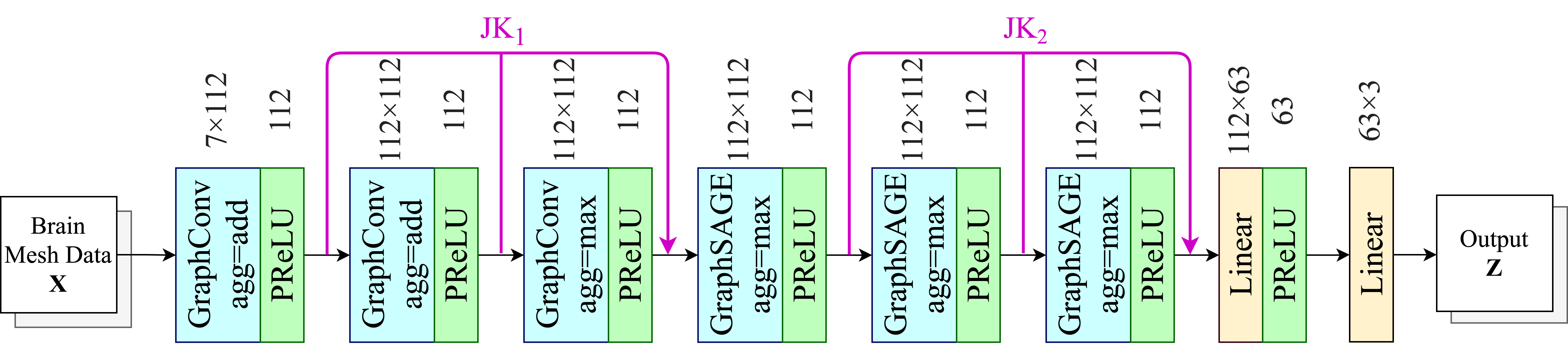}
    \captionsetup{justification=centering}
    \caption{}
    \label{fig:best_config_2}
    \end{subfigure}
    \begin{subfigure}[b]{0.3\textwidth}
    \includegraphics[width=\linewidth]{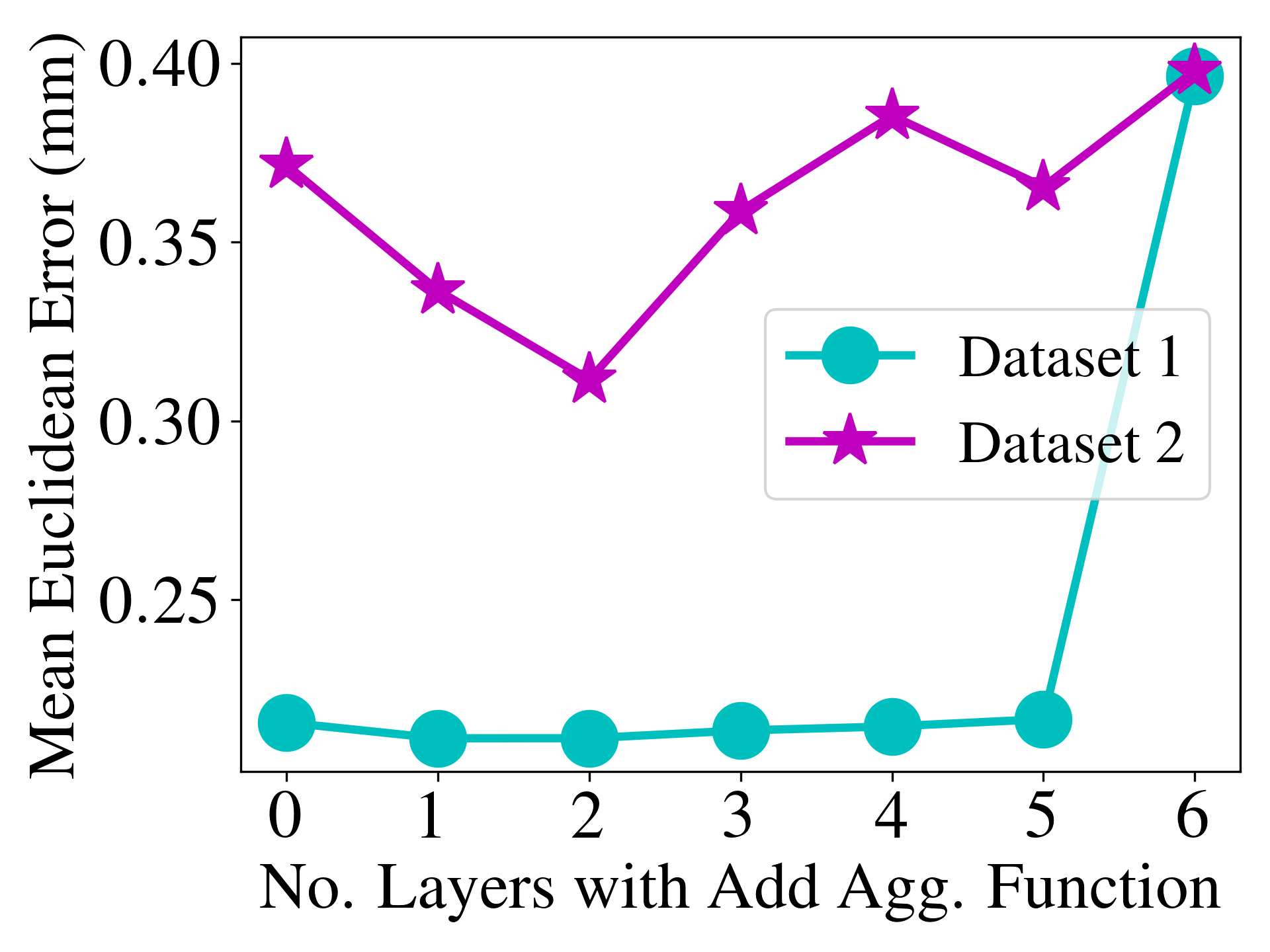}
    \captionsetup{justification=centering}
    \caption{}
    \label{fig:ablation_agg}
    \end{subfigure} \hspace{0.1\textwidth}
    \begin{subfigure}[b]{0.3\textwidth}
    \includegraphics[width=\linewidth]{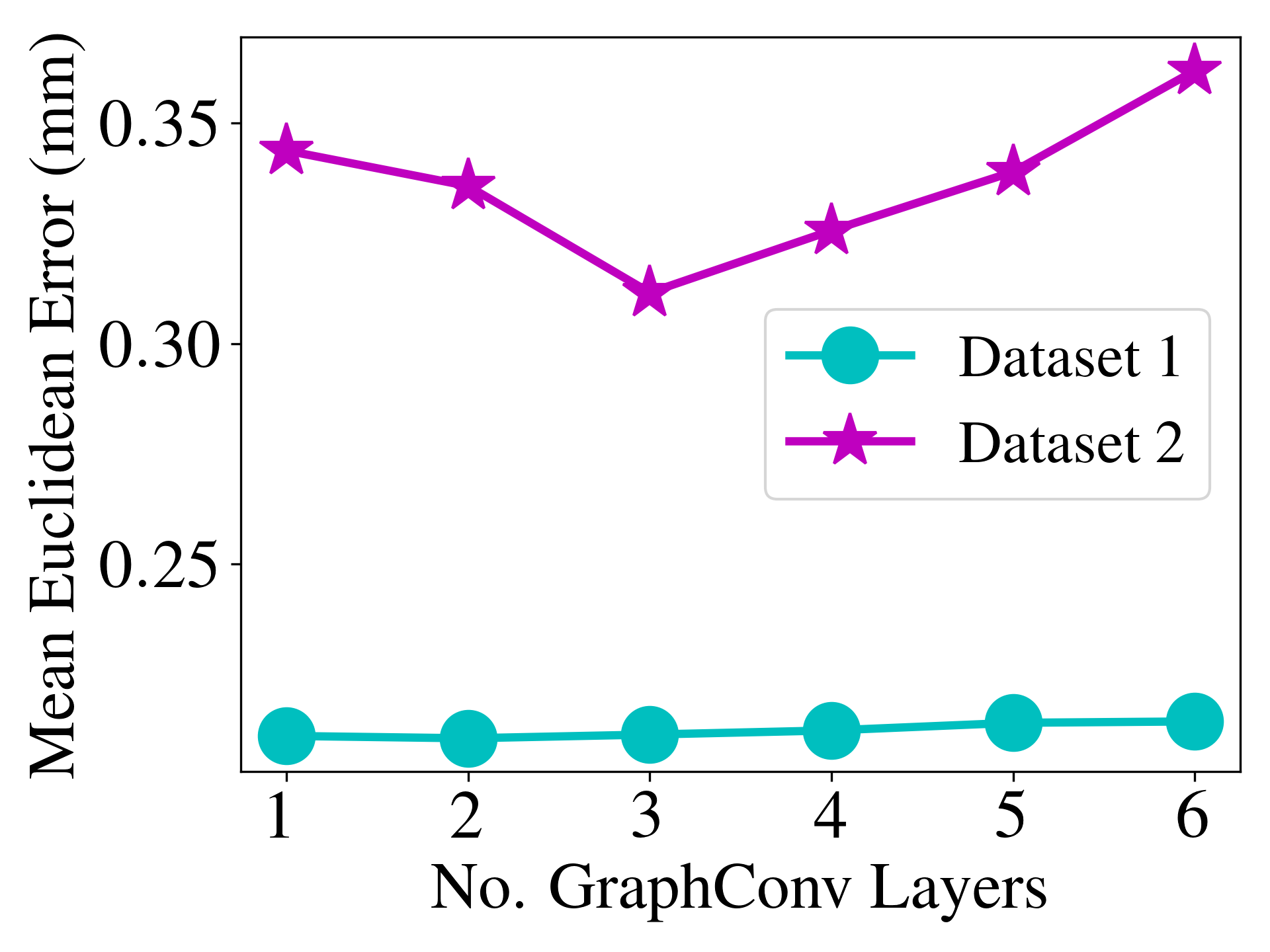}
    \captionsetup{justification=centering}
    \caption{}
    \label{fig:ablation_gc}
    \end{subfigure} 
\caption{Best-performing PhysGNN models on (a) Dataset 1, and (b) Dataset 2. The effect of (c) consecutively replacing the maximum with the addition aggregation function in a PhysGNN model consisting of 3 GraphConv layers followed by 3 GraphSAGE layers, and (d) consecutively replacing GraphSAGE with GraphConv layers while using the best-performing aggregation function combination determined from Figure \ref{fig:ablation_agg}. Agg. stands for aggregation.}
\label{best_configs_and_ablation}
\end{figure}

% \begin{figure}[t]
%  \begin{tabular}{m{0.6\hsize}m{0.25\hsize}}
%  \begin{subfigure}{0.6\textwidth}
%     \includegraphics[width=\linewidth]{images/PhysGNN_best_d1.png}
%     \captionsetup{justification=centering}
%     \caption{}
%     \label{fig:best_config_1}
%     \end{subfigure}

%     \begin{subfigure}{0.6\textwidth}
%     \includegraphics[width=\linewidth]{images/PhysGNN_best_d2.png}
%     \captionsetup{justification=centering}
%     \caption{}
%     \label{fig:best_config_2}
%     \end{subfigure}
% & 
%     \begin{subfigure}{0.25\textwidth}
%     \includegraphics[width=\linewidth]{images/ablation_agg.png}
%     \captionsetup{justification=centering}
%     \caption{}
%     \label{fig:ablation_agg}
%     \end{subfigure} 
    
%     \begin{subfigure}{0.25\textwidth}
%     \includegraphics[width=\linewidth]{images/ablation_gc.png}
%     \captionsetup{justification=centering}
%     \caption{}
%     \label{fig:ablation_gc}
%     \end{subfigure} 
%     \end{tabular}

% \caption{Best-performing PhysGNN configuration on (a) Dataset 1, and (b) Dataset 2. The effect of (c) consecutively replacing the maximum with the addition aggregation function in a PhysGNN model consisting of 3 GraphConv layers followed by 3 GraphSAGE layers, and (d) consecutively replacing GraphSAGE with GraphConv layers while using the best-performing aggregation function combination determined from Figure \ref{fig:ablation_agg}. Agg. stands for aggregation.}
% \label{best_configs_and_ablation}
% \end{figure}

To find the best-performing PhysGNN model on different datasets, we assessed the performance of various PhysGNN models resulted from different combinations of GNN layers and aggregation functions. First, to find the best-performing aggregation function combination, we started with a PhysGNN model consisting of three GraphConv layers followed by three GraphSAGE layers with all six layers using the maximum aggregation function, which we consecutively replaced with the addition aggregation function. Afterwards, we initialized a PhysGNN model with one GraphConv layer followed by five GraphSAGE layers and replaced GraphSAGE with GraphConv layers consecutively while using the best-performing combination of aggregation functions. 

Initializing the PhysGNN model with three GraphConv layers followed by three GraphSAGE layers, consecutive replacement of maximum with addition aggregation function for two iterations reduced the mean validation Euclidean loss of Dataset 2 more noticeably, while such replacement for five iterations only marginally changed the performance of PhysGNN on the validation set of Dataset 1, as shown in Figure \ref{fig:ablation_agg}. Similarly, consecutive replacement of GraphSAGE with GraphConv layers while using the best-performing combination of aggregation functions resulted in a more pronounced decrease in the validation loss of Dataset 2---where using three GraphConv layers yielded the best performance--- while caused over-smoothing beyond a certain number for both datasets, as depicted in Figure \ref{fig:ablation_gc}. 

These observations are justified by realizing that nodal displacements in Dataset 1 are resulted from applying forces to only one node per simulation, contrary to nodal displacements in Dataset 2, which occur from multiple prescribed force nodes per simulation. Therefore, for Dataset 2, not only is using the summation aggregation function a more reasonable choice, but also weighting the impact of multiple force load points on their neighboring nodes by multiplying their message with the inverse of the nodal distance is indeed expected to produce more accurate results. The later observation sheds light on the importance of incorporating nodal distance and proper formulation of edge weights to effectively capture tissue deformation. 

It is worthy to note that the best-performing PhysGNN model on Dataset 2, depicted in Figure \ref{fig:best_config_2}, resulted in a validation and test mean Euclidean error of $0.2114$ mm and $0.2058$ mm, respectively, for Dataset 1, which is only marginally greater than the errors resulted from the best-performing model for Dataset 1, depicted in Figure \ref{fig:best_config_1}. Accordingly, the PhysGNN model depicted in Figure \ref{fig:best_config_2} may be regarded as robust for both datasets. 

% These observations are justified by realizing that nodal displacements in Dataset 1 are resulted from applying forces to only one node, contrary to nodal displacements in Dataset 2, which occur from multiple prescribed force nodes. Consequently, while using the maximum aggregation function in different GNN layers would be a more reasonable choice when training PhysGNN with Dataset 1, summing prescribed forces applied to the nodes within the same neighbourhood results in more accurate tissue displacement predictions when training PhysGNN with Dataset 2.

%  However, while weighting neighbouring nodes' messages with edge weights using GraphConv layers decreased loss value in both datasets for the first few iterations, it caused over-smoothing beyond a certain number. 

\subsection{Limitations and Future Directions}

% One of the limitations of this study was that comparing our results to other studies could only be done on empirical grounds due to either lack of accessibility to other studies' data and/or implementation, or not having sufficient computational power to use their data/models. However, our work presents itself as a great benchmark for future studies for having all the data and implementation publicly accessible and only requiring a Google Colab Pro server for reproducing the results. 
One of the limitations of this study is that comparing our method against other works was done on empirical grounds due to either lack of accessibility to other studies' data and/or implementation, or not having sufficient computational power to use their data/models. However, our work presents itself as a great benchmark for future studies as our data and implementations are publicly accessible and reproducing our results only requires a Google Colab Pro server.

Another limitation of this work is using the FEA results as baseline, which is dependent on the quality of MRI image segmentation, and the selected soft tissue constitutive laws that may not hold true for different patients \citep{tonutti2017machine}. Moreover, tetrahedral elements are prone to undergoing volumetric locking for almost incompressible materials, such as soft tissue, and should not be used to simulate tissue deformation \citep{joldes2019suite}. To alleviate these problems, Meshless Total Lagrangian Explicit Dynamics (MTLED) algorithms are emerging as a solution to better simulate soft tissue deformation, which can also be trained with empirical results \citep{miller2019biomechanical, joldes2019suite}. As PhysGNN may also be trained with the results of Lagrangian meshless methods, the presented framework is foreseen to yield highly accurate and computationally efficient data-driven deformable volumes in the future.
\section{Conclusion}
The purpose of this work was to introduce a novel data-driven framework, PhysGNN, to combat the limitations of using the FEM for approximating tissue deformation---and especially craniotomy-induced brain shift---in neurosurgical settings. Through our empirical results, we demonstrated the effectiveness of PhysGNN in accurately predicting tissue deformation. Compared to similar studies, our method is competitive with data-driven SOTA approaches for capturing tissue deformation while promising enhanced computational feasibility, especially for large and high-quality meshes.

\begin{ack}
The authors would like to thank the reviewers for their helpful and constructive feedback. This work was supported by the Natural Sciences and Engineering Research Council (NSERC) of Canada.
\end{ack}

%%%%%%%%%%%%%%%%%%%%%%%%%%%%%%%%%%%%%%%%%%%%%%%%%%%%%%%%%%%%

\bibliography{main}
\bibliographystyle{plainnat}

\section*{Checklist}

%%% BEGIN INSTRUCTIONS %%%
% The checklist follows the references.  Please
% read the checklist guidelines carefully for information on how to answer these
% questions.  For each question, change the default \answerTODO{} to \answerYes{},
% \answerNo{}, or \answerNA{}.  You are strongly encouraged to include a {\bf
% justification to your answer}, either by referencing the appropriate section of
% your paper or providing a brief inline description.  For example:
% \begin{itemize}
%   \item Did you include the license to the code and datasets? \answerYes{See Section~\ref{gen_inst}.}
%   \item Did you include the license to the code and datasets? \answerNo{The code and the data are proprietary.}
%   \item Did you include the license to the code and datasets? \answerNA{}
% \end{itemize}
% Please do not modify the questions and only use the provided macros for your
% answers.  Note that the Checklist section does not count towards the page
% limit.  In your paper, please delete this instructions block and only keep the
% Checklist section heading above along with the questions/answers below.
%%% END INSTRUCTIONS %%%

\begin{enumerate}

\item For all authors...
\begin{enumerate}
  \item Do the main claims made in the abstract and introduction accurately reflect the paper's contributions and scope?
    \answerYes{}
  \item Did you describe the limitations of your work?
    \answerYes{}
  \item Did you discuss any potential negative societal impacts of your work?
    \answerNA{}
  \item Have you read the ethics review guidelines and ensured that your paper conforms to them?
    \answerYes{}
\end{enumerate}

\item If you are including theoretical results...
\begin{enumerate}
  \item Did you state the full set of assumptions of all theoretical results?
    \answerNA{}
        \item Did you include complete proofs of all theoretical results?
    \answerNA{}
\end{enumerate}

\item If you ran experiments...
\begin{enumerate}
  \item Did you include the code, data, and instructions needed to reproduce the main experimental results (either in the supplemental material or as a URL)?
    \answerYes{}
  \item Did you specify all the training details (e.g., data splits, hyperparameters, how they were chosen)?
    \answerYes{}
        \item Did you report error bars (e.g., with respect to the random seed after running experiments multiple times)?
    \answerYes{} The standard deviation of the errors are reported in Tables \ref{tab:results_data} and \ref{tab:max_vals}. Regarding the error bars with respect to the random seed, the random seed was initialized to 1 for all experiments to allow for reproducible results. 
        \item Did you include the total amount of compute and the type of resources used (e.g., type of GPUs, internal cluster, or cloud provider)?
    \answerYes{}
\end{enumerate}

\item If you are using existing assets (e.g., code, data, models) or curating/releasing new assets...
\begin{enumerate}
  \item If your work uses existing assets, did you cite the creators?
    \answerYes{} The code used towards generating the datasets was adapted from the examples provided in the Gibbon \citep{moerman2018gibbon} documentation. Additionally, one of the python functions used towards creating the PyTorch Geometric datasets was adapted from the code provided in \citep{you2019position}, which is referred to in the ReadMe file of the supplementary material that has been published in the GitHub repository. The surface mesh models taken from \citep{tonutti2017machine}, which are based on MRI scans in \citep{madhavan2009rembrandt} are also cited in the paper.
  \item Did you mention the license of the assets?
    \answerYes{}{} The licences of the aforementioned code assets are included in the license document that is published in the GitHub repository.
  \item Did you include any new assets either in the supplemental material or as a URL?
    \answerYes{} The code used towards generating the datasets, as well as all the PhysGNN models have been included in the supplementary material, which are also available at the public GitHub repository. 
  \item Did you discuss whether and how consent was obtained from people whose data you're using/curating?
    \answerYes{} All the data used from other sources---i.e. the surface mesh of the brain model taken from \cite{tonutti2017machine} that was generated from the MRI images taken from the Repository of Molecular Brain Neoplasia Data \cite{madhavan2009rembrandt}---have been cited in the paper. 
  \item Did you discuss whether the data you are using/curating contains personally identifiable information or offensive content?
    \answerNA{}
\end{enumerate}

\item If you used crowdsourcing or conducted research with human subjects...
\begin{enumerate}
  \item Did you include the full text of instructions given to participants and screenshots, if applicable?
    \answerNA{}
  \item Did you describe any potential participant risks, with links to Institutional Review Board (IRB) approvals, if applicable?
    \answerNA{}
  \item Did you include the estimated hourly wage paid to participants and the total amount spent on participant compensation?
    \answerNA{}
\end{enumerate}

\end{enumerate}

\appendix

\section{Finite Element Volume Mesh}

% You can have as much text here as you want. The main body must be at most $8$ pages long.
% For the final version, one more page can be added.
% If you want, you can use an appendix like this one, even using the one-column format.

Table \ref{tab:mesh_info} provides information on the finite element (FE) volume mesh that was used for generating Datasets 1 and 2. 

\begin{table}[h]
\caption{Finite element volume mesh statistics.}
\label{tab:mesh_info}
\centering
    \begin{tabular}{cc}
    \toprule
    Attribute & Value \\
    \midrule
     Mesh points  & 9118 \\
     Mesh tetrahedra    & 55927 \\
     Mesh faces & 112524 \\
     Mesh faces on exterior boundary & 1340 \\
     Mesh faces on input facets & 2992 \\
  Mesh edges on input segments & 4488 \\
  Steiner points inside domain & 7618 \\
  \toprule
    \end{tabular}
\end{table}

\section{Infrastructure Settings}
The FE simulations in our study were carried out on quad-core Intel i7 @ 2.9 GHz CPU, while different PhysGNN models were trained on a Google Colab Pro server with 23GB of RAM and one NVIDIA P-100 GPU with 16 GB of video RAM.

\section{Additional Results}

The table below represents the median of absolute error in the $x$, $y$, and $z$ directions denoted as $\delta x$, $\delta y$, and $\delta z$ respectively, Euclidean error, and absolute position error. 

% \begin{table}[h]
%     % \centering
%     \begin{tabular}{cccccc}
%     {>{\centering\arraybackslash}m{0.57in}
%     >{\centering\arraybackslash}m{0.52in}
%     >{\centering\arraybackslash}m{0.52in}
%     >{\centering\arraybackslash}m{0.52in}
%     >{\centering\arraybackslash}m{0.56in}
%     >{\centering\arraybackslash}m{0.45in}
%     }
%     \toprule
%      Dataset & ($\delta x$) (mm) & ($\delta y$) (mm) & ($\delta z$) (mm) & Median Euclidean Error (mm) &
%      Median Absolute Position Error (mm) \\
%     \toprule
%          &  \\
%          & 
%     \end{tabular}
%     \caption{Caption}
%     \label{tab:my_label}
% \end{table}

\begin{table}[h]
\caption{Median of the evaluation metrics resulted from the best-performing PhysGNN models on Datasets 1 and 2.}
\label{tab:results_median}
\centering
% \begin{small}
    \begin{tabular}
    {>{\centering\arraybackslash}m{0.6in}
    >{\centering\arraybackslash}m{0.6in}
    >{\centering\arraybackslash}m{0.6in}
    >{\centering\arraybackslash}m{0.6in}
    >{\centering\arraybackslash}m{0.6in}
    >{\centering\arraybackslash}m{0.8in}
    }
    \toprule
     Dataset & Median Absolute Error ($\delta x$) (mm) & Median Absolute Error ($\delta y$) (mm) & Median Absolute Error ($\delta z$) (mm) & Median Euclidean Error (mm) &
     Median Absolute Position Error (mm)  \\
    \toprule
    Dataset 1 Validation & 0.0199 & 0.0228 & 0.0221 & 0.0531 & 0.0303 \\
    \hline
    Dataset 1 Test & 0.0200 & 0.0227 & 0.0224 & 0.0537 & 0.0303 \\
    \toprule
    Dataset 2 Validation & 0.0436 & 0.0537 & 0.0538 & 0.1295 & 0.0656 \\
    \hline
    Dataset 2 Test & 0.0426 & 0.0534 & 0.0529 & 0.1285 & 0.0646 \\
    \toprule
    \end{tabular}
% \end{small}
\end{table}

% \section{Material Properties}

% Table \ref{tab:mat_properties} provides information on the tumour and healthy brain tissue material parameters used for carrying out FE analysis.

% \begin{table}[h]
% \caption{Material properties of different parts of the brain \citep{joldes2010real, wittek2010patient}.}
% \label{tab:mat_properties}
% \centering
%     \begin{tabular}
%     {cccc}
%     \toprule  
%      Dataset & Density (kg/$\text{m}^3)$ & Young's Modulus ($\text{Pa}$) & Poisson Ratio  \\
%     \midrule
%     Healthy Tissue & 1000 & 3000 & 0.49\\
%     Tumour & 1000 & 7500 & 0.49 \\
%     \bottomrule
%     \end{tabular}
% \end{table}

% \section{Features and Outputs of PhysGNN}

% Table \ref{tab:input_output} provides a summary of the inputs to and outputs from PhysGNN, as discussed in Section 4.2.

% \begin{table}[h]
% \caption{PhysGNN features and outputs.}
% \label{tab:input_output}
% \centering
%     \begin{tabular}{cc}
%     \toprule
%     Features ($\mathbf{X}$) & Output ($\mathbf{Z}$)\\
%     \toprule
%      $F_x, F_y, F_z, F_\rho, F_\theta, F_\phi,$ Physical Property & $\delta x$, $\delta y$, $\delta z$ \\
%      \toprule 
%     \end{tabular}
% \end{table}

% \begin{equation*}
% \small
% \frac{\text{Young's Modulus}_{\text{healthy tissue}}}{\text{Young's Modulus}_{\text{tumour tissue}}} = \frac{3000 \text{ Pa}}{7500 \text{ Pa}} = 0.4.    
% \end{equation*}

% \begin{equation*}
% \begin{cases}
%     0.4, & \text{if the node belongs to tumour tissue, and} \\
%     1, & \text{if the node belongs to healthy brain tissue,}
% \end{cases}
% \end{equation*}

\end{document}